\documentclass[aps,twocolumn,superscriptaddress,floatfix,showpacs,amsmath,longbibliography,nofootinbib,amssymb]{revtex4-1}
\usepackage[colorlinks=true,citecolor=blue,linkcolor=blue]{hyperref}
\usepackage[usenames]{color}
\usepackage{txfonts}
\usepackage{subfigure}
\usepackage{dcolumn}
\usepackage{mathrsfs}
\usepackage{bm}
\usepackage{amsmath,amssymb,epsfig,float,graphics}
\DeclareMathOperator{\sech}{sech}

\begin{document}
\baselineskip=0.45 cm

\title{Analogue Hawking Radiation and Sine-Gordon Soliton in a Superconducting Circuit}

\author{Zehua Tian}
\email{tianzh@ustc.edu.cn}
\affiliation{CAS Key Laboratory of Microscale Magnetic Resonance and Department of Modern Physics, University of Science and Technology of China, Hefei 230026, China}

\affiliation{Hefei National Laboratory for Physical Sciences at the Microscale, University of Science and Technology of China, Hefei 230026, China}

\affiliation{Synergetic Innovation Center of Quantum Information and Quantum Physics, University of Science and Technology of China, Hefei 230026, China}

\author{Jiangfeng Du}
\email{djf@ustc.edu.cn}
\affiliation{CAS Key Laboratory of Microscale Magnetic Resonance and Department of Modern Physics, University of Science and Technology of China, Hefei 230026, China}
\affiliation{Synergetic Innovation Center of Quantum Information and Quantum Physics, University of Science and Technology of China, Hefei 230026, China}
\affiliation{Hefei National Laboratory for Physical Sciences at the Microscale, University of Science and Technology of China, Hefei 230026, China}

\begin{abstract}

We propose the use of a waveguide-like transmission line based on direct-current superconducting quantum interference devices (dc-SQUID) and study the sine-Gordon (SG) equation which characterises the dynamical behavior of the superconducting phase in this transmission line.  Guided by the duality between black holes in Jackiw-Teitelboim (JT) dilaton gravity and solitons in sine-Gordon field theory, we show how to, in our setup, realize $1+1$ dimensional black holes as solitons of the sine-Gordon equation. We also study the analogue Hawking radiation in terms of the quantum soliton evaporation, and analyze its feasibility within current circuit quantum electrodynamics (cQED) technology. Our results may not only facilitate experimentally understanding the relation between Jackiw-Teitelboim dilaton gravity and sine-Gordon field theory, but also pave a new way, in principle, for the exploration of analogue quantum gravitational effects.

\end{abstract}

\baselineskip=0.45 cm
\maketitle
\newpage
\textit{Introduction.}--- The sine-Gordon (SG) equation, which was first introduced in 1862 by Edmond Bour when studying the surfaces of constant negative curvature, is a nonlinear hyperbolic partial differential equation involving the d$^\prime$Alembert operator and the sine of the unknown function. This equation relates to several fields in mathematics touching upon differential geometry and nonlinear partial differential equation. In addition, it also has been fruitfully applied to the description of a diverse range of areas of physics and biology, such as one-dimensional crystal dislocation theory \cite{Frank205, Kochend1950, Seeger1951, Seeger1953}, fluxon dynamics in Josephson transition line \cite{Scott1970, PhysRevA.18.1652}, excitation of phonon modes \cite{PhysRevE.64.056608}, condensation of charge density waves \cite{1126-6708-2005-08-023, PhysRevD.77.023523, PhysRevD.77.023523, PhysRevLett.36.432}, and DNA-promoter dynamics \cite{PhysRevA.44.5292, PhysRevE.66.016614, PhysRevLett.99.154101}. In particular, $1+1$ dimensional black holes of JT dilaton gravity can be realized as solitons of the SG equation \cite{GEGENBERG1997274}. The SG theory thus may provide an alternative classical description of the simplest two dimensional gravity theory.

Black hole is believed to exist in binary systems as well as the center of most galaxies \cite{black-holes}. Besides, black hole involves to 
some fundamental problems and the relevant resolution will likely provide important clues about the interface between quantum mechanics and gravity. However, since typical stellar size black holes are cold and young and are far away from us, 
it is almost impossible to settle the relevant issues, such as demonstrating Hawking radiation, through direct astrophysical black hole observations. Therefore, the investigations on black hole have remained mostly theoretical to date. To render the relevant physics accessible to an experimental investigation, the so called ``analogue gravity" \cite{BarcelA2011}, in which the relevant features of quantum fields in curved spacetime can be reproduced analogously, has been constructed. Recently, a lot of ``analogous gravity" experiments involving many nascent yet fast-growing fields, such as Bose-Einstein-Condensates \cite{PhysRevLett.85.4643, PhysRevLett.105.240401} and ion trap \cite{PhysRevLett.104.250403, 1367-2630-13-4-045008}, have been proposed to observe the analogue Hawking radiation. 

The duality between JT black holes and SG solitons might be used to shed light on the field theory origin of black holes and the dynamical source of black hole entropy \cite{GEGENBERG1997274, PhysRevD.54.6206}. In particular, this intriguing connection opens a new avenue for the analog of black hole as well as the observation of the relevant quantum gravity effects, e.g., Hawking radiation.
In this Letter, we propose using a superconducting electrical circuit configuration to study the nonlinear SG field theory that has deep connections with the JT dilaton gravity. We investigate the realization of $1+1$ dimensional black holes as solitons of the SG equation
and analyze how to detect the analogue Hawking radiation in terms of the quantum soliton evaporation. Note that our scenario is based on the well established cQED technology \cite{cite-key}, which could offer a natural arena for testing fundamentals of quantum mechanics and implementing quantum field theory (QFT) concepts \cite{RevModPhys.84.1} due to the antastic controllability and scalability. Moreover, our setup is quite similar to the superconducting coplanar waveguide (CPW) \cite{PhysRevA.82.052509, PhysRevLett.103.147003} and the SQUID array transmission line \cite{2007ApPhL91h3509C, SUID-array} (see more details below), which have already been constructed experimentally with parameters near those required in our setup to observe the Hawking effect \cite{2007ApPhL91h3509C, SUID-array, DCE, Josephson-metamaterial}. Our results therefore can in principle be achievable within the current cQED technology.

\begin{figure}[ht]
\centering
\includegraphics[width=0.48\textwidth]{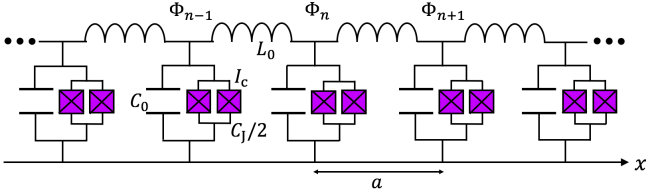}
\caption{(Color online) Circuit diagram for a coplanar waveguide-like transmission line. The inductance for each inductor and the capacitance for each capacitor are assumed to be $L_0$ and $C_0$, respectively. Each SQUID element consists of two identical tunnel Josephson junctions with critical current $I_\text{c}$ and capacitance $\frac{1}{2}C_\text{J}$. The length of all the cells is the constant $a$. The circuit is characterized by the dynamical fluxes $\Phi_n$.}\label{fig1}
\end{figure}

\textit{Physical model.}--- As shown in Fig. \ref{fig1}, we propose to use a coplanar transmission line similar to the CPW in Ref. \cite{PhysRevA.82.052509, PhysRevLett.103.147003}. However, in our configuration each capacitor is parallel with an identical SQUID.
Note that the added SQUIDs can provide a nonlinear potential to the Lagrangian of fluxes in the CPW, and plays a very crucial role in the simulation of black hole (see details in the following). In this work, we assume that each SQUID is symmetric, i.e., its two parallel Josephson junctions (JJs) have the equal critical current $I_\text{c}$ and capacitances $\frac{1}{2}C_\text{J}$. Besides, the geometric size of SQUID loop is assumed to be small enough such that its self-inductance is negligible compared to its kinetic inductance. As a consequence of that, each SQUID can be referred to as an effective JJ with a junction capacitance $C_\mathrm{J}$ and a tunable Josephson energy $E_\text{J}(\Phi^\text{J}_\text{ext})=2E_\text{J}\cos\bigg(\pi\frac{\Phi^\text{J}_\text{ext}}{\Phi_0}\bigg)$ \cite{PhysRevLett.103.147003}. Here $\Phi_0=h/2e$ is the magnetic flux quantum, $E_\text{J}=\frac{\Phi_0I_\text{c}}{2\pi}$ is the Josephson energy, and $\Phi^\text{J}_\text{ext}= BA_\text{S}$ is the flux dropping through the SQUID loop with effective area $A_\text{S}$, and the applied magnetic field $B$.

With the quantum-network theory \cite{PhysRevA.29.1419}, the Lagrangian corresponding to our circuit configuration reads
\begin{eqnarray}\label{Lagrangian-S}
\nonumber
\mathscr{L}&=&\sum^{N}_{n=1}\bigg[\frac{1}{2}C_0\big(\dot{\Phi}_n\big)^2-\frac{(\Phi_{n+1}-\Phi_n)^2}{2L_0}+\frac{1}{2}C_\text{J}\big(\dot{\Phi}_n\big)^2
\\
&&+E_\text{J}(\Phi^\text{J}_\text{ext})\cos\bigg(2\pi\frac{\Phi_n}{\Phi_0}\bigg)\bigg],
\end{eqnarray}
where $\Phi_n$ is the node flux. We restrict ourselves to the macroscopic SQUID junctions in the phase regime, i.e., when the Josephson energy is so big compared to the charging energy, $E_\text{J}(\Phi^\text{J}_\text{ext})\gg(2e)^2/2C_\text{J}$, and oscillations in the phase across the SQUID could satisfy $\frac{\Phi_n}{\Phi_0}\ll1$. The small amplitude condition allows us to study the Lagrangian above by linearizing the Josephson cosine terms. In this regard, let us note we have expanded the cosine terms in \eqref{Lagrangian-S} up to the second order in $\Phi^2_n$ to investigate the analog cosmological particle creation \cite{PhysRevD.95.125003}. However, here we will learn the whole nonlinear terms without any expansion. It is these cosine terms that produce a sine term for the Klien-Gordon equation of the superconducting phase (shown in \eqref{field2}), which plays a key role in the analog of black hole. Furthermore, we assume that the wavelength $\lambda$ for the flux is much longer than the dimensions of a single unit cell of the chain, i.e., $a/\lambda\ll1$. It implies that the continuum approximation is valid in our scenario. We can thus replace the discrete $n$ by a continuous position $x$ along the line and replace the finite difference in the Lagrangian by their continuous counterparts to the first order in $(a/\lambda)$, i.e., $\Phi_n-\Phi_{n-1}\approx\,a\frac{\partial\Phi}{\partial\,x}+o(a^2)$. Finally, in the continuum limit, the Lagrangian in Eq. \eqref{Lagrangian-S} can be rewritten as
\begin{eqnarray}\label{L2}
\nonumber
\mathscr{L}&=&\frac{C}{2}\int\,dx\bigg[\bigg(\frac{\partial{\Phi}}{\partial\,t}\bigg)^2-\frac{a^2}{L_0C}\bigg(\frac{\partial\Phi}{\partial\,x}\bigg)^2+\frac{E_\text{J}(\Phi^\text{J}_\text{ext})}{C}
\\
&&\times\cos\bigg(\frac{2\pi\Phi}{\Phi_0}\bigg)\bigg],
\end{eqnarray}
where $C=C_0+C_\text{J}$. Through variation with respect to $\phi=\frac{2\pi}{\Phi_0}\Phi$ and its derivative, 
we find the superconducting phase $\phi$ satisfies $1+1$ dimensional SG equation, 
\begin{eqnarray}\label{field2}
\frac{\partial^2\phi}{\partial\,t^2}-c^2\frac{\partial^2\phi}{\partial\,x^2}+m^2\sin\phi=0.
\end{eqnarray}
Here $c=a/\sqrt{L_0C}$ is the velocity of propagation, which in practice is well below the vacuum speed of light $c_0$, and $m=\sqrt{4\pi^2E_\text{J}(\Phi^\text{J}_\text{ext})/C\Phi^2_0}$ can be considered as the effective mass. For simplification, hereafter we will assume $c=1$ similar to the conventional natural units.

It has long been known that the solutions of the SG equation \eqref{field2} determine Riemannian geometries with constant negative curvature $-2m^2$ \cite{Bullough}. The corresponding metric characterizing this manifold is given by the line-element,
\begin{eqnarray}\label{line-element1}
ds^2=\sin^2\bigg(\frac{\phi}{2}\bigg)dt^2+\cos^2\bigg(\frac{\phi}{2}\bigg)dx^2, 
\end{eqnarray}
where the angle $\phi$ describes the embedding of the manifold into a three dimensional Eucildean space \cite{Bullough}. However, 
$ds^2$ in \eqref{line-element1} does not remain Lorentz invariant under general coordinate transformations and thus does not lead to a Schwarzschild-like metric in the future analysis. In order to obtain the Lorentz invariant one, we perform a Wick rotation $t\rightarrow\,it$ following Ref. \cite{GEGENBERG1997274}. Besides, adopting the new variables $\tau=mt$ and $\xi=mx$, we can obtain the elliptic SG equation:
\begin{eqnarray}\label{field3}
\frac{\partial^2\phi}{\partial\tau^2}+\frac{\partial^2\phi}{\partial\xi^2}=\sin\phi,
\end{eqnarray}
whose corresponding Lorentz invariant metric is given by the line-element,
\begin{eqnarray}\label{line-element2}
ds^2=-\sin^2\bigg(\frac{\phi}{2}\bigg)d\tau^2+\cos^2\bigg(\frac{\phi}{2}\bigg)d\xi^2.
\end{eqnarray}

\textit{Jackiw-Teitelboim gravity and sine-Gordon soliton.}---The JT dilaton gravity is a theory of gravity in $1+1$ dimensions in which a scalar field (the dilaton) is non-minimally coupled to the spacetime metric \cite{QTG}. 
In this theory,  the spacetime $M_2$ has a Lorenzian metric $g_{\mu\nu}$ and its Ricci curvature is a negative constant, i.e., $R(g)=-2m^2$ ($g$ is the determinant of the metric $g_{\mu\nu}$). The corresponding action for JT gravity is 
\begin{eqnarray}\label{JT-action}
I_\text{JT}[\psi, g]=\frac{1}{2G}\int_{M_2}dx^2\sqrt{|g|}\psi(R+2m^2),
\end{eqnarray}
where $G$ is the gravitational coupling constant that is dimensionless in two dimensional spacetime, and $\psi$ is the dilaton field.  
As shown in Ref. \cite{GEGENBERG1997274}, sufficient conditions that this functional be stationary under arbitrary variations of the dilaton and metric fields are, respectively, 
\begin{eqnarray}\label{D-M-equation}
R+2m^2=0,
\\
(\nabla_\mu\nabla_\nu-m^2g_{\mu\nu})\psi=0.
\end{eqnarray}
We can solve the above equations with the metric \eqref{line-element2}. It is straightforward to show that this metric has constant negative curvature $R=-2m^2$ if and only if $\phi$ satisfies the Euclidean SG equation in \eqref{field3}. Moreover, in this case the dilaton $\psi$
satisfies the linearized equation,
\begin{eqnarray}\label{dilaton-equation}
(\partial^2_\tau+\partial^2_\xi)\psi=\cos(\phi)\psi.
\end{eqnarray}
Note that the dilaton, which generates the Killing vectors (i.e., symmetries) of the black hole metric, also maps solutions of the SG equation onto other solutions \cite{GEGENBERG1997274}. That is to say, if $\phi$ and $\psi$ obey Eqs. \eqref{field3} and \eqref{dilaton-equation}, respectively, then the field $\phi^\prime=\phi+\epsilon\psi$, also solves Eq. \eqref{field3} to the first order in $\epsilon$. 

Although all the solutions of JT gravity are locally diffeomorphic to two-dimensional anti-DeSitter spacetime, one may obtain distinct global solutions, some of which display many of the attributes of black holes \cite{PhysRevD.51.1781, PhysRevD.47.4438}. As an example, we shall 
now demonstrate that the 1-soliton solution of the elliptic SG equation \eqref{field3} determines a black hole metric.
The 1-soliton solution of the elliptic SG equation can be written as 
\begin{eqnarray}\label{solution1}
\phi(\tau,\xi)=4\arctan{\exp[\pm\gamma(\xi-\beta_s\tau)]},
\end{eqnarray}
where $\gamma=(1+\beta_s^2)^{-1/2}$ and the constant $\beta_s$ is a ``spectral parameter" satisfying $0<\beta_s<1$. The solution with the 
$+$ sign in the exponent is the 1-soliton solution, while the opposite sign is the anti-soliton solution. By substituting Eq. \eqref{solution1} into the Lorentzian metric \eqref{line-element2}, the latter can be reduced to
\begin{eqnarray}\label{line-element3}
ds^2=ds^2_\text{1-sol}=-\sech^2\rho\,d\tau^2+\tanh^2\rho\,d\xi^2,
\end{eqnarray}
with $\rho=\gamma(\xi-\beta_s\tau)$. Following Ref. \cite{GEGENBERG1997274}, we can perform a successive coordinate transformations to obtain the  Schwarzschild form of \eqref{line-element3},
\begin{eqnarray}\label{metric1}
ds^2=(\beta^2_s-r^2)d\mathcal{T}^2-(\beta^2_s-r^2)^{-1}dr^2.
\end{eqnarray}
Here the definitions $d\mathcal{T}=d\tau-\beta_s\frac{\tanh^2\rho}{\gamma(\sech^2\rho-\beta^2_s\tanh^2\rho)}d\rho$ and $r=\frac{1}{\gamma}\sech\rho$ have bee used. Eq. \eqref{metric1} is the metric of a JT black hole with event horizon at $r_\text{H}=\beta_s$. It is actually a dimensionally truncated three dimensional BTZ black hole \cite{PhysRevLett.69.1849, PhysRevD.48.1506, PhysRevD.57.1068}.

In order to display the geometries of black hole, we introduce the analogue tortoise coordinate. Impose $r^\ast(r)=\frac{1}{2\beta_s}\ln\bigg(\frac{\beta_s+r}{\beta_s-r}\bigg)$ so that $dr^\ast=(\beta_s-r^2)^{-1}dr$, then Eq. \eqref{metric1} can be rewritten as 
\begin{eqnarray}\label{metric2}
ds^2=[\beta^2_s-r^2(r^\ast)][d\mathcal{T}^2-(dr^\ast)^2].
\end{eqnarray}
This coordinates also have a singularity at $r=\beta_s$ and span only the exterior of black hole. Besides, when $r$ approaches to the event horizon $\beta_s$, $r^\ast$ goes to $+\infty$, while far away from the black hole (i.e., $r=0$), $r^\ast$ goes to zero. By introducing the Kruskal lightcone coordinates $(u,v)$ which are associated with the lightcone coordinates $\tilde{u}=\mathcal{T}-r^\ast$, $\tilde{v}=\mathcal{T}+r^\ast$ with $u=e^{\beta_s\tilde{u}}/\beta_s$, $v=e^{-\beta_s\tilde{v}}/\beta_s$, we can rewrite the metric above as 
\begin{eqnarray}\label{metric3}
ds^2=[\beta_s+r(\tilde{u}, \tilde{v})]^2dudv.
\end{eqnarray}
Note that here the metric \eqref{metric3} is regular at $r=\beta_s$. Therefore, the singularity $r=\beta_s$ occurring in both the elliptic SG-soliton metric and the Schwarzschild one is, in fact, a coordinate singularity, which can be removed by a coordinate transformation. The Kruskal coordinates span the entire spacetime.


\textit{Hawking radiation.}---General relativity and quantum field theory predict that in the background of black hole particles can be created from vacuum fluctuation near the black hole horizon. The temperature of the radiation spectrum is proportional to the surface gravity of the black hole. This is known as Hawking effect \cite{Hawking, Hawking1975}.

Consider a massless scalar field in the background of black hole, we can solve its equation of motion in the above two coordinates \eqref{metric2} and \eqref{metric3}, and obtain two sets of complete basises. Based on the quantum field theory in curved spacetime \cite{Birrell1982ix, Mukhanov:1122230}, we can expand arbitrary quantum field with these two basises as 
\begin{eqnarray}
\nonumber
\Psi&=&\int^\infty_0\frac{d\Omega}{2\sqrt{\pi\Omega}}\bigg[e^{-i\Omega\tilde{u}}\hat{b}_\Omega+e^{i\Omega\tilde{u}}\hat{b}^\dagger_\Omega\bigg]+\text{left moving}
\\  \nonumber
&=&\int^\infty_0\frac{d\omega}{2\sqrt{\pi\omega}}\bigg[e^{-i\omega\,u}\hat{a}_\omega+e^{i\omega\,u}\hat{a}^\dagger_\omega\bigg]+\text{left moving},
\end{eqnarray}
where the left moving part is given by terms wighted by $e^{\pm\Omega\tilde{v}}$ or $e^{\pm\omega\,v}$ in the mode expansion. The two sets of creation and annihilation operators specify two different vacuum states, i.e., Boulware vacuum $\hat{b}_\Omega|0_\text{B}\rangle=0$ and Kruskal vacuum $\hat{a}_\omega|0_\text{K}\rangle=0$. Note that the Boulware vacuum, in fact, is singular at the horizon. However, the Kruskal coordinates span the entire spacetime and thus the Kruskal vacuum is regular on the horizon and corresponds to true physical vacuum in the presence of the black hole. According to the Bogoliubov transformation between the above two modes, one can obtain 
\begin{eqnarray}\label{spectrum}
\langle\hat{N}_\Omega\rangle=\langle0_\text{K}|\hat{b}^\dagger_\Omega\hat{b}_\Omega|0_\text{K}\rangle=\bigg[\exp\bigg(\frac{2\pi\Omega}{\beta_s}\bigg)-1\bigg]^{-1},
\end{eqnarray}
which is a thermal spectrum with the corresponding temperature $T_\text{H}=\beta_s/2\pi$. It means that seen from the remote observer moving with the soliton tail, the Kruskal vacuum is not vacuum anymore. That is to say, particles are emitted from black hole.

Alternatively, we can also obtain analogue Hawking radiation in the SG field theory by analyzing the weak perturbation of field \cite{PhysRevB.90.224509}, i.e., by assuming $\phi\simeq\phi_s+\phi_1$. Here $\phi_s$ describes the classical solution to Eq. \eqref{field3}, and $\phi_1\ll\phi_s$ is the perturbation which satisfies
\begin{eqnarray}\label{perturbation-equation}
[\partial^2_\tau+\partial^2_\xi]\phi_1-\cos(\phi_s)\phi_1=0,
\end{eqnarray}
to the first order of $\phi_1$. Following Ref. \cite{2399-6528-2-5-055016}, we can obtain the same result as that shown in \eqref{spectrum}.

Note that the Hawking temperature $T_\text{H}=\beta_s/2\pi$ depends on the velocity of soliton, $\beta_s$. This dependence of the Hawking radiations on the translation velocity is peculiar of soliton dynamics \cite{PhysRevD.58.084025} and it is related to the structure of the spectral parameter in the inverse scattering transform \cite{PhysRevLett.31.125, PhysRevLett.30.1262}. 
Besides, the elliptic SG soliton moves with the velocity $\beta_s$ and the frequency $\Omega$ seen by an observer at rest with respect to the soliton thus should contain a Doppler shift \cite{2399-6528-2-5-055016}. As a consequence of that, the Hawking temperature in the laboratory frame reads 
\begin{eqnarray}\label{temperature}
T_\text{H}=\frac{\beta_s}{2\pi}\sqrt{\frac{1-\beta_s}{1+\beta_s}},
\end{eqnarray}
which can be reduced to $T_\text{H}\simeq\frac{\beta_s}{2\pi}(1-\beta_s)$ in the small $\beta_s$ limit.

\begin{figure}[hbt]
\centering
\subfigure[]{\includegraphics[width=0.38\textwidth]{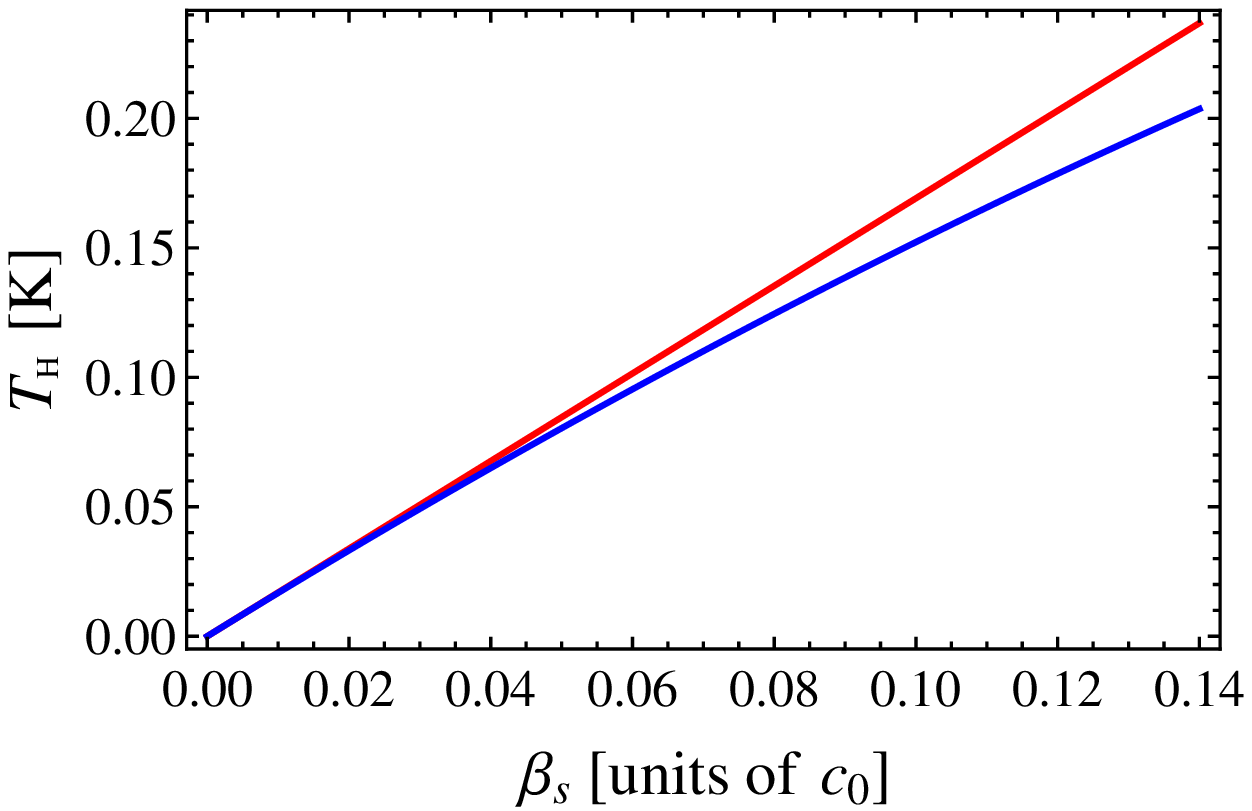}}
\subfigure[]{\includegraphics[width=0.38\textwidth]{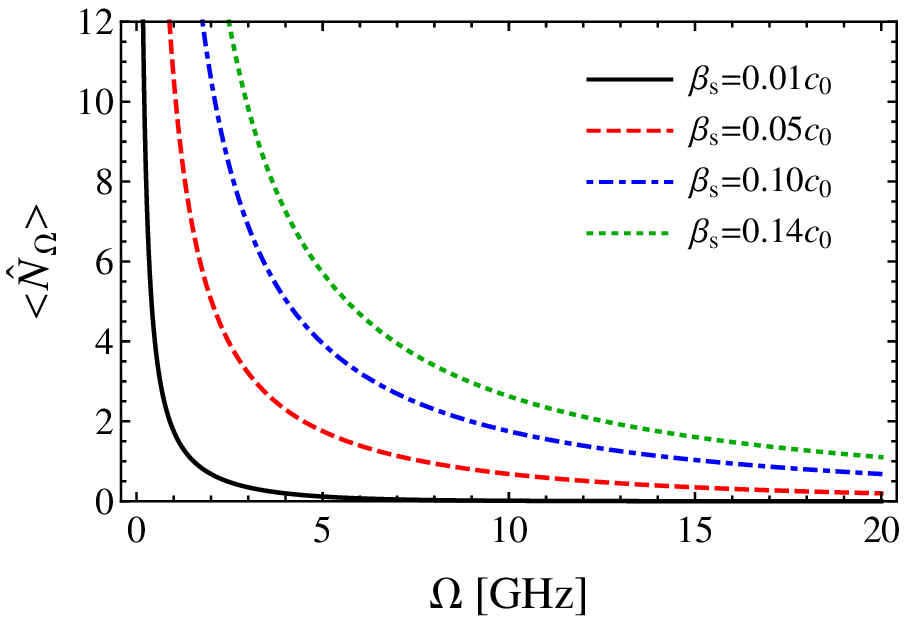}}
\caption{(Color online) (a) The temperature of analogue Hawking radiation observed in comoving frame (red solid line) and in laboratory frame (blue solid line); (b) the radiation spectrum of Eq. \eqref{spectrum} as a function of particle frequency seen from the comoving frame. Note that the radiation spectrum seen from the laboratory, which is not shown, just has very slightly quantitive difference.}\label{fig2}
\end{figure} 

\textit{Experimental implementation.}---To realize the proposed experiment, we import a signal $\phi_s$ plus a linear weak perturbation $\phi_1\ll\phi_s$ into our setup. The signal $\phi_s$ behaves as SG soliton and plays the role of black hole, while $\phi_1$ can be considered as a quantum perturbation of the black hole metric and induce the analogue Hawking radiation that we aim to observe \cite{2399-6528-2-5-055016}. The soliton velocity 
$\beta_s$ can be set by preparation of the signal and must not be higher than the unbiased transmission line propagation velocity $c$. Therefore, the Hawking radiation should satisfy $T_\text{H}\leq\,c/2\pi$ in the comoving frame and the radiation power is given by \cite{PhysRevLett.95.031301, PhysRevA.62.052104, PhysRevLett.103.087004}
\begin{eqnarray}
\frac{dE}{d\mathcal{T}}=\frac{\pi}{12\hbar}(k_bT_\text{H})^2.
\end{eqnarray}
In order to observe the quantum fluctuation, i.e., the analogue Hawking radiation, a frequency-tunable, single-shot photon detection at the end of the transmission line should be prepared. For example, we may choose a superconducting phase qubit as the detector and detect the microwave photon based on the recently proposed technologies \cite{PhysRevLett.101.240401, PhysRevLett.102.173602, PhysRevLett.107.217401, PhysRevLett.117.140503}. Alternatively, as done in Ref. \cite{DCE}, we may also measure the four quadrature voltages of the upper and lower sidebands which can be used to demonstrate the voltage-voltage correlations of output predicted by Hawking radiation theory.

We will choose the relevant parameters for each element similar to Refs. \cite{PhysRevA.82.052509, PhysRevLett.103.147003, 2007ApPhL91h3509C, SUID-array, DCE, Josephson-metamaterial, PhysRevLett.103.087004} and estimate the analogue Hawking radiation in our configuration. For the JJ, we assume its critical current and effective junction capacitance are respectively $I_\text{c}=2\,\mu\mathrm{A}$ and $C_\text{J}=1.2\,\mathrm{fF}$, and the plasma frequency is thus $\omega_s=2\pi\sqrt{E_\text{J}/\Phi^2_0C_\text{J}}\simeq2.25\times10^{12}\,\mathrm{Hz}$. Besides, the capacitance to ground is chosen as $C_0=0.8\,\mathrm{fF}$, the inductance and the length of the single unit cell of our setup are respectively assumed to be $L_0=0.01\,\mathrm{nH}$ and $a=6\,\mu\mathrm{m}$. In this case, the propagation velocity is $c\simeq0.14c_0$ where $c_0$ is the velocity of light in the vacuum. In Fig. \ref{fig2}, we plot the Hawking temperature and the spectrum of Hawking radiation. It is shown that the effective temperature is proportional to the velocity of the soliton, and the number of created particle decreases as the increase of the particle frequency. Let us note that the temperature could be as high as a few $\mathrm{mK}$ which can be a factor of 10 larger than the ambient temperature set by a dilution refrigerator. Therefore, this effect should be visible above the background noise.

\textit{Discussions and Conclusions.}---The relevant analysis can be extended to a more general case, i.e., constructing SG coordinates for a black hole with N-soliton \cite{PhysRevD.58.124010}. In addition, the effective mass $m$ in Eq. \eqref{field2} actually can be adjusted by means of a suitable strongly inhomogeneous external magnetic flux bias along a waveguide-like transmission line, and thus is time- and space-dependent. This property allows us to investigate, in our current setup, the relevant issues in Ref. \cite{PhysRevLett.99.154101}, i.e., the SG model with a variable mass involving a nonuniform Josephson junction and DNA-promoter dynamics, and so on. Note that the parameters and pulse shapes in our paper were chosen as an example that our setup is feasible, which should not be considered as the only available configuration. In fact, it is possible to improve and optimize these values in terms of both performance and fabrication of this proposal. On the other hand, properly engineering the transmission line could effectively reduce the background noise from the unwanted coupling and make the detection of analogue Hawking radiation more effective.

In summary, we have provided a recipe to build up a quantum simulator of black hole physics based on the superconducting circuit. In our setup, the superconducting phase satisfies the SG equation. Therefore, due to the duality between black holes in JT dilaton gravity and solitons in SG field theory, the solutions of the SG equation can be used to realize the constant curvature two dimensional black holes.
Our results showed that the proposed device works in the quantum region and allows us to observe the analogue Hawking radiation.


\begin{acknowledgments}
This work was supported by the National Key R\&D Program of China (Grant No. 2018YFA0306600), the CAS (Grants No. GJJSTD20170001 and No. QYZDY-SSW-SLH004), and Anhui Initiative in Quantum Information Technologies (Grant No. AHY050000).
\end{acknowledgments}


\bibliography{Soliton-BH}

\end{document}